\documentclass[runningheads]{llncs}
\usepackage[T1]{fontenc}
\usepackage{amssymb}
\usepackage{amsmath}
\usepackage{amsfonts}
\usepackage{float}
\usepackage{appendix}
\usepackage{enumitem}
\usepackage{mathtools}
\setenumerate[1]{label=(\arabic*)}
\usepackage{enumitem}
\usepackage{algpseudocode}
\usepackage{algorithm}

\begin{document}
\title{New Quantum-Safe Versions of Decisional Diffie–Hellman Assumption  in the General Linear Group and Their Applications}
\subtitle{ Two Key-agreements}
\author{Abdelhaliem Babiker}
\institute{Department of Basic Eng. Sciences, College of Engineering, Imam Abdulrahman Bin Faisal University- Dammam, Saudi Arabia\\
\email{aababiker@iau.edu.sa; haliem.abbas@gmail.com} }

\titlerunning{DH-WE Key-agmnt and RSA-Resemble.}
\maketitle

\begin{abstract}Diffie-Hellman key-agreement and \textsl{RSA} cryptosystem are \\ widely used to provide security in internet protocols. But both of the two algorithms are totally breakable using Shor's algorithms. This paper proposes two connected matrix-based key-agreements: (a) \textit{Diffie-Hellman Key-Agreement with Errors} and (b) \textit{ RSA-Resemble Key-agreement}, which, respectively, bear resemblance to Diffie-Hellman key-agreement and \textsl{RSA} cryptosystem and thereby they gain some of the well-known security characteristics of these two  algorithms, but without being subject to Shor's algorithms attacks. That is, the new schemes avoid the direct reliance on the hardness of Discrete Logarithm and Integer Factoring problems which are solvable by Shor's algorithms. 
The paper introduces a new family of quantum-safe hardness assumptions which consist of taking noisy powers of binary matrices. The new assumptions are derived from \textsl{Decisional Diffie-Hellman (DDH)} assumption in the general linear group $GL(n,2)$ by introducing random noise into a quadruple  similar to that which define the \textsl{DDH} assumption in $GL(n,2)$. Thereby we make certain that the resulting quadruple is secure against Shor's algorithm attack and any other \textsl{DLP}-based attack. Thence, the resulting assumptions,  are used as  basis for the two  key-agreement schemes. We prove that these key-agreements are secure ---in key indistinguishability notion--- under  the new assumptions.

\end{abstract}
\section{Introduction}
\numberwithin{equation}{section}
The mid-seventies of the last century had witnessed  the birth of the \textit{Public Key Cryptography}, a concept introduced by W. Diffie and M. Hellman in their seminal  work \cite{diffie1976new} in which they proposed what will become later known as Diffie-Hellman key-agreement, a widely used  algorithm for establishing session keys in internet protocols \cite{adrian2015imperfect}.  In 1978, two years after the paper of W. Diffie and M. Hellman, \textsl{RSA}  public key cryptosystem  for encryption and digital signature had been invented  by Ron Rivest, Adi Shamir, and Len Adleman \cite{rivest1978method}, which has also become widely used to provide privacy and authenticity for many applications \cite{boneh1999twenty}.\\ \indent  Diffie-Hellman key-agreement and \textsl{RSA} cryptosystem are based on the hardness of two different but related mathematical problems. The first is based on the hardness of the \textit{Discrete Logarithm Problem (DLP)}, and the second is based on hardness of the \textit{Integer Factorization Problem}. In 1994 Peter Shor \cite{shor1994algorithms} invented algorithms that use quantum computers to break  these two problems.\\ \indent  Recent advancements in the field of quantum computers are approaching towards developing large-scale quantum computers, and it is expected that large-scale quantum computers will be built in the foreseeable future. If this happens, then the Diffie-Hellman key-exchange, the \textsl{RSA} cryptosystem, and many other public key cryptosystems will be broken using Shor's algorithms which utilizes quantum computers for solving the the problems upon which these cryptosystems relies. This serious threat posed by the quantum computers has created  great interest in developing cryptograms that are based on  hardness assumptions which are secure against both of classical and quantum computers \cite{257451}.

\indent This paper proposes a family of new hardness assumptions which are used as basis for two connected key-agreement algorithms  that, respectively, avoid the reliance  on the Discrete Logarithm Problem and the Integer Factorization problem while retaining similar structure to  Diffie-Hellman key-agreement and \textsl{RSA} cryptosystem. The two algorithms have common underlying hardness assumption, which  is hardness of finding certain information from noisy binary matrix. More precisely, \textit{a problem of finding information about a secret random  matrix  $M \in \mathbb{F}_2 ^{n \times n} $ given 
\begin{equation}
\label{error_correction_eqn}
A=M^{\alpha} \oplus E,
\end{equation}
where $E \in \mathbb{F}_2 ^{n \times n} $ is a secret random (error) matrix and $\alpha$ is  nonnegative integer.}
 The operator $\oplus$ is the bitwise XOR operator. Thus, the matrix $A$ is result of bitwise XOR between two secret binary matrices, and the question is to find information about $M$; particularly the secret number $\alpha$ or the multiplicative order of $M$.  \\

In this introduction we give general overview of the two schemes and their security. More detailed description is given in the subsequent sections. \\

\subsection{Two Connected Key-agreements Using Noisy Powers of Binary Matrices}
  For the purpose of this paper, Diffie-Hellman  key-agreement, denoted (\textsl{DH}) key-agreement, in which two parties, conventionally named Alice and Bob, agree on a shared secret key $K$ is described as follows. 
 \begin{enumerate}
 \item Alice generates random nonsingular matrix $M \in \mathbb{F}_2 ^{ n \times n}$ and sends ${(M, B=M^{\alpha})}$ to Bob, for secret random integer ${\alpha}$.
 \item Bob selects secret random integer $\beta$,  computes $X=B^{\beta}= M^{\alpha \beta}$, and ${Y=M^{\beta}}$ , then sends $Y$ to Alice. 
 \item Alice computes $Y^{\alpha}=M^{\beta \alpha}$. 
 \end{enumerate}
The shared secret key is $ K = M^{ \alpha \beta}=M^{  \beta \alpha}$.\\
 
Generally, any adversary who is able to solve the scalar-\textsl{DLP} can easily break this scheme using the fact that for every eigenvalue $\lambda$  of the matrix $M$, the numbers $\lambda ^{\alpha}$ and $\lambda ^{\beta}$ are, respectively, eigenvalues for $M^{\alpha}$ and $M^{\beta}$, and hence the pairs $(\lambda, \lambda ^{\alpha})$ and $(\lambda, \lambda ^{\beta})$ define DLPs that gives $\alpha$ and $\beta$. Thus, this matrix-based classical \textsl{DH} is totally breakable using Shor's algorithm for the \textsl{DLP}.\\

To prevent any \textsl{DLP}-based attack on this classical \textsl{DH} scheme  we introduce the following new scheme which involves random error matrices added to the  messages exchanged between the two parties of the \textsl{DH} key-agreement described above. 
\subsection*{(I) Diffie-Hellman Key-Agreement with Errors}
A new scheme \textit{Diffie-Hellman Key-Agreement with Errors (DH-WE)} is defined by adding random (error) matrices to each of the matrices $M$, $M^{\alpha}$, and $M^{\beta}$ in the original \textsl{DH} key-agreement as follows. 
 \begin{enumerate}
 \item Alice sends $(A = M\oplus E_0, B = M^{\alpha} \oplus E_1 ,S)$ to Bob, for secret binary matrices $M$, $E_0$, and $E_1$, secret random integer ${\alpha}$, and random binary matrix $S$.
 
\item  Next, Bob selects secret random number $\beta$, computes $k_B= B^{\beta}_j$-- the $j$th column of the matrix $B^{\beta}$, generates secret random matrix $R_2$, and computes \[ {Y=A^{\beta} \oplus R_2S},\]
 then send $Y$ to Alice.
 
[Actually, $Y=A^{\beta} \oplus  R_2 S = M^{\beta} \oplus E_2 $, for some unknown matrix $E_2$.]
\item Alice computes the secret key $k_A = Y^{\alpha}_j$ -- the $j$th column of the matrix $Y^{\alpha}$.  
 \end{enumerate}
The shared secret key is $k=k_A=k_B$. The \textsl{DH-WE} scheme is specified by the \textit{tuple} \[\Pi_{M,\alpha\beta}= (A,B,S,Y,k).\] 
We may imagine the \textsl{DH-WE} scheme as a classical \textsl{DH} scheme in which the eavesdropping adversary between Alice and Bob receives noisy (corrupted) messages instead of the real messages exchanged between the two parties (in particular $A$, $B$, and $Y$ instead of $M$, $M^{\alpha}$, and $M^{\beta}$, respectively), so she must detect and get-rid of the noise before solving the relevant \textsl{DLP} for $\alpha$ or $\beta$. Hence, if error correction problem (\ref{error_correction_eqn}) is hard, then it is hard for the eavesdropping adversary to solve any of the two relevant \textsl{DLP}s and thereby the proposed scheme is secure against any DLP-based attack. 
 \subsection*{(II) \textsl{RSA}-Resemble Key-agreement}
 The \textsl{RSA} public key cryptosystem is used as public key encryption scheme for key establishment and also used as digital signature algorithm. In both cases there are two keys: the public key $(e,n)$, and the private key $d$, where 
\[n=pq,\text{ } d= e^{-1} \mod \phi, \text{ and  } \phi=(p-1)(q-1),\]
for two  secret large prime numbers $p$ and $q$. \\
\indent In the encryption scheme, the public key is used for encryption and the private key for the decryption. A message $m$, which is an integer less than $n$, is encrypted using the public key $(e,n)$ as 
$ c = m^e \mod n $ and the private key is used by the legitimate receiver to decrypt the ciphertext and obtain the plaintext message as $ m = c^d \mod n $. The signature scheme works in a similar way in which the private key is used for signature generation and the public key is used for verification of the signature. 

\indent Security of the \textsl{RSA} cryptosystem is based on the secrecy of the number $\phi$ which is  product of the two secret numbers $(p-1)$ and $(q-1)$. Thus, the \textsl{RSA} is based on the hardness of factoring the modulus $n$ into its prime factors $p$ and $q$. \\

As mentioned before, the problem of factoring  is breakable using Shor's algorithm. \\

\indent We proposes new key agreement algorithm that bears some similarity with \textsl{RSA} cryptosystem. The sender in the proposed scheme encrypts a binary matrix $X$ using public exponent $e$ as $Y= X^e$, and the receiver uses private number $d$ to compute $Y^{d}=X^{ed}$, where the number $d$ is, as in the \textsl{RSA}, the multiplicative inverse of $e$ modulo secret number $\phi$. That is 
\[d=e^{-1} \mod \phi.\]
However, unlike the \textsl{RSA}, the modulus $\phi$ in the proposed scheme is not  the \textit{Euler's Phi Function} of a composite number $n=pq$. Instead, the modulus $\phi$ is chosen to be the multiplicative order of a secret  ${ n \times n}$ binary matrix $M$. In other words, the number $\phi$ satisfies 
\[M^{\phi} = I_n, \text{ where }  I_n \text{ is the identity matrix}. \]
We chose $e$ and $d$ such that 
\[ M^{ed}= M^{q \phi +1}=M , \text{ for some integer } q.\]
\indent Security of the new scheme is based on the secrecy of the matrix $M$. That is, since the matrix $M$ is secret, its multiplicative order $\phi$ is also secret; thus it is hard to find the private number $d$ given the number $e$. \\
\indent We give here general description of the key-agreement scheme, and more detailed description will be given in  Section 5. In this scheme Alice and Bob agree on a shared secret key as follows.  
 \begin{enumerate}
 \item Alice sends $(A = M\oplus E_0, B = M^{\alpha} \oplus E_1 ,e,j)$ to Bob, for secret binary matrices $M$ (with multiplicative order $\phi$), $E_0$, and $E_1$, secret random integer ${\alpha}$ and random integer $e$. 
 
\item  Bob selects two secret random numbers $\theta$ and $\vartheta$ computes $X= A^{\theta}B^{\vartheta}$, sets $k_B= X_j$ -- the $j$th column of the matrix $X$. 
\item Next, Bob  computes $Y=X^{e}$ then sends $Y$ to Alice.
 
[In fact, $Y= M^{\beta} \oplus E_2 $, for some integer $\beta$ and some unknown matrix $E_2$.]
\item Alice computes the secret key $k_A = Y^{d}_j$ -- the $j$th column of the matrix $Y^d$, where $d= e^{-1} \mod \phi$.  
 \end{enumerate}
The shared secret key is $k=k_A=k_B$. The \textsl{RSA}-Resemble key-agreement scheme is specified by the \textit{tuple} \[\Pi_{M,e,d}= (A,B,e,Y,k).\]

\subsection{Security of the Two Schemes}
We first consider the key-recovery attack on each of the two schemes, then we mention briefly the main notion of security of each of the two schemes, that is the notion of key indistinguishability under  certain assumption. \\  
 \indent It is easy to see that security of both of the two schemes relies on the secrecy the matrix $M$. The two schemes can, respectively, be specified as
 \[\Pi_{M,\alpha\beta}=(A,B,S,Y,k) \text{ and } \Pi_{M,e,d}=(A,B,e,Y,k),\]
 where 
 \[A=M \oplus E_0,\text{ } B= M^{\alpha} \oplus E_1,\text{and } Y = M^{\beta} \oplus E_2, \] 
for the random secret matrices $M$, $E_0$, $E_1$, and $E_2$  and random secret integers $\alpha$ and $\beta$. 
\subsubsection{Security Against Key Recovery Attack} Key-recovery attack on the first scheme, ($\Pi_{M,\alpha\beta} $), requires finding one of the secret exponents $\alpha$ or $\beta$ which are obtainable only through the secret matrix $M$. The attack is defined by the  problem: \\  $ \textit{ Given the tuple } \Pi_{M,\alpha\beta}, \textit{ find }  k_B = B^{\beta}_j \textit { or } k_A = Y^{\alpha}_j$,  \textit{for some publicly known column index  j}.\\
 To find one of the secret numbers $\alpha$ or $\beta$ one must solve one of the matrix DLPs $(M,M^{\alpha})$ or $(M,M^{\beta})$. However, the matrix $M$ is secret as well as $M^{\alpha}$ and $M^{\beta}$. All of what the attacker has is only noisy powers of the secret matrix $M$, that is $A$, $B$, and $Y$. \\
\indent Likewise, security of the \textsl{RSA}-Resemble key-agreement against key-recovery attack is defined by the problem:\\
 \textit{ Given } the tuple $\Pi_{M,e,d}$, \textit{find} $k_A = Y^{d}_j \textit { or } k_B = X_j$, for some publicly known $j$,\textit{ where} $X = A^{\theta}B^{\vartheta}$, \textit{for the secret numbers $\theta$ and $\vartheta$}.\\ Thus, the key-recovery attack on this scheme requires either finding the secret matrix $X$ which is obviously hard since $X = A^{\theta}B^{\vartheta}$, or, alternatively, finding  the private key $d = e^{-1} \mod \phi $ which is hard to compute without the secret number $\phi$ ---the multiplicative order the secret matrix $M$. \\
 
\indent Thus, security of each of the two schemes against key-recovery based on the secrecy of the matrix $M$ which is hard to be recovered from the noisy matrices 
\[A = M \oplus E_0 \text{, } B= M^{\alpha} \oplus E_1,\text{ and } Y = M^{\beta} \oplus E_2,  \]
where $E_0$, $E_1$, and $E_2$ are random secret matrices from some error distribution $ \mathcal{E} \subset \mathbb{F}_2 ^{n \times n}$ and $\alpha$ and $\beta$ are secret integers.
\subsubsection{Security Against Key Distinguishability Attack} Full key-recovery attack is weak security notion. Therefore, a stronger security notion of key indistinguishability from the random is used to define the security of each of the  the two schemes. 
We introduce a new family of hardness assumptions similar to \textsl{Decisional Diffie-Hellman (DDH)} assumption in the general linear group $GL(n,2)$. For the first scheme, the assumption is defined by adding random noise to the quadruple of the matrices that defines the \textsl{DDH} assumption in $GL(n,2)$. Thus, unlike the \textsl{DDH} assumption, the new assumption is not subject to \textsl{DLP}-based attacks. The second scheme is based on a similar assumption with slight differences. Full detailed description of these assumptions is given in Section 3. \\

\indent The rest of this paper is organized as follows. We lay down the basic foundations for both of the two schemes in Section 2, namely the matrix $M$ and its properties. In Section 3 we present  some  definitions and elucidations and we articulate  our assumptions. Then we present in Sections 4 and 5 consecutively the two key-agreements in full details. Finally, we conclude in Section 6. 
 
\section{Preliminaries}
In the context of this paper all matrices are square matrices over the binary field  $\mathbb{F}_2$. Arithmetic of matrix multiplication and matrix addition is carried out in $\mathbb{F}_2$. We use the operator $+$ interchangeably with the operator $\oplus$  to denote addition in $\mathbb{F}_2$ or bitwise XOR according to the  context. \\
\paragraph {Let  
\begin{equation}
\label{M}
M = F^{-1}GF,
\end{equation}
where $F,G \in \mathbb{F}_2 ^{n \times n}$ , such that \\
\[ F_{i,j}=0 , \text{  for }  1\leq i \leq l\text{ and some fixed } j, \]  and 
\[ 
G = 
\begin{bmatrix}
D_{l\times l} &  \\
 & C_{m\times m} \\
\end{bmatrix},\hspace{2pt} 
\text{with   } D^2 =I_l \hspace{1pt}, \text{ where } I_l \text{ is } l\times l \text{ indentity matrix}. \]
}
\indent Let $r$ be the multiplicative order of the matrix $C$, such that $r$ is odd. Notice that the matrix $D$ has the multiplicative order 2. Thus, the matrix $G$ has multiplicative order $\phi$ of $2r$. That is, 
\[
G^{\phi} = 
\begin{bmatrix}
D^{2r} &  \\
 & C^{2r} \\
\end{bmatrix}
=
\begin{bmatrix}
I_l &  \\
 & I_m \\
\end{bmatrix}
= I_n,\vspace{2pt}\text {where } I_n \text { is the identity matrix}.
\]
Hence, the multiplicative order of the matrix $M$ is $\phi$.

\subsection{Construction of the Matrix $M$}
It is easy to construct matrix $M$ as defined in \eqref{M}. One way to do that is as follows. \\
\textbf{\textit{The Matrix}} $F$:
Generate nonsingular matrix $F \in \mathbb{F}^{n \times n}$, then select some column that contains at least $l$ zeros. Let us assume that this column is the $j$th column. Sort the rows of $F$ such that the first $l$ elements of the $j$th column are all zeros. Thus, we get 
\[F_{i,j}=0 , \text { for } 1\leq i \leq l \text{ and this particular  $j$th column}.\] 
Next, we compute $F^{-1}$ using Gaussian elimination.\\
\textbf{\textit{The Matrix}} $D$:  We can choose the following matrix
\[ D =  \begin{bmatrix}
d & &\\
& \ddots & \\
 & & d
\end{bmatrix}
 , \text { where } d= \begin{bmatrix}
 1 & 1\\
 0 & 1
\end{bmatrix}.  \]
 or any matrix similar to $D$. For example, a matrix $EDE^{-1}$ for some $l \times l$ invertible matrix $E$. \\
\textbf{\textit{The Matrix}} $C$: We Construct the matrix $C$ such that $C$ has multiplicative order $r$. The matrix $C$ can be constructed as \\
$$ C =  \begin{bmatrix}
c_1 & &\\
& \ddots & \\
 & & c_t
\end{bmatrix}, 
 $$ where  $c_i$, for $1\leq i \leq t$, is a companion matrix of which a minimal polynomial $ p_i(x)$ $\in \mathbb{F}_2[x]$  has multiplicative order $r_i$.
 Thus, the multiplicative order of $C$ is the least common multiple of $r_1$,\dots, $r_t$. That is, $r = lcm(r_1,...,r_t)$. \\
 
\indent Next, let us define the matrix $Q$ as $$Q = (I_n + M^{r}).$$
Since $M=F^{-1}GF$, it follows that 
 \[Q=(I_n +M^r)=(I_n +F^{-1} G^rF )=F^{-1}(I_n +G^r)F,\] 
 where $G^r = 
\begin{bmatrix}
D^r &  \\
 & C^r \\
\end{bmatrix} =
\begin{bmatrix}
D &  \\
 & I_m \\
\end{bmatrix}$. \vspace{6pt}\\
Notice that $r$ is the multiplicative order of the matrix $C$ and thus $C^{r}=I_m$. The number $r$ is odd and the multiplicative order of $D$ is 2, hence $D^{r}$=$D$. \vspace{6pt}\\
Now, $I_n+G^{r} = I_n +
\begin{bmatrix}
D &  \\
 & I_m \\
\end{bmatrix} =
 \begin{bmatrix}
I_l+D &  \\
 & I_m+I_m \\
\end{bmatrix}=
\begin{bmatrix}
I_l +D &  \\
 & 0_{m \times m} \\
 \end{bmatrix} $. \vspace{6pt}\\
Thus, 
$$Q=(I_n + M^{r})=F^{-1}(I_n +G^r)F= F^{-1} \begin{bmatrix}
I_l+D &  \\
 & 0_{m \times m} \\
 \end{bmatrix} F.$$ 
\begin{remark}
Since $F_{i,j} = 0$, for $1 \leq i \leq l$ and the certain $j$th column, it follows that the $j$th column in the matrix $Q$ is zero column. That is, $Q_{i,j}=0$, for $1 \leq i \leq n$, or simply $Q_j=0$. 
\end{remark}
\begin{theorem}
\label{thrm1}
Let $\alpha$, $\theta$ and $\vartheta$ be  nonnegative integers, and let $R$, $\Sigma _1 $ and $ \Sigma _2 $ be matrices in $\mathbb{F}_2 ^{n \times n }$. Then, 
\begin{align}
\label{mq_q}
M^{\theta}Q=QM^{\theta} &=Q\\
\label{m_alpha_rq_rq}
(M{^\alpha} +RQ)^{\theta} &= M^{\alpha\theta}+\Sigma Q\\
\label{m_alpha_beta_rq_rq}
(M^{\theta}+\Sigma _1Q)(M^{\vartheta} +\Sigma _2 Q)&= M^{\theta+\vartheta}+ \Sigma _3 Q
\end{align}
for some unknown matrices $\Sigma$, $\Sigma _3$ $\in$  $\mathbb{F}_2 ^{n \times n }$.
\end{theorem}
\begin{proof} 
\begin{enumerate}[label=(\alph*)]
\item  $M^{\theta}Q=QM^{\theta} =Q$. \\
It is easy to see that $QM^{\theta}=M^{\theta}Q$. The result follows directly from definition of the matrix $Q$.\\
Next, we have 
$Q=(I_n + M^{r})=F^{-1}(I_n +G^r)= F^{-1} \begin{bmatrix}
I_l+D &  \\
 & 0_{m \times m} \\
 \end{bmatrix} F$. \\
 On the other hand 
 \begin{align*}
 QM&=(I_n+M^r)M\\
              		   &=(I_n+{F^{-1}G}^r F)F^{-1}GF\\
        		   &= F^{-1}(I_n+G^r)FF^{-1}GF\\
        		   &= F^{-1}(I_n+G^r)GF.
 \end{align*}
 Since $I_n + G^r = \begin{bmatrix}
I_l +D &  \\
 & 0_{m \times m} \\
 \end{bmatrix}$ and $G=\begin{bmatrix}
D &  \\
 & C \\
 \end{bmatrix}$,
 \begin{align*}
 QM &= F^{-1}\left(I_n+G^r\right)GF\\
 &=F^{-1} \begin{bmatrix}
I_l +D &  \\
 & 0_{m \times m} \\
 \end{bmatrix} \begin{bmatrix}
D &  \\
 & C \\
 \end{bmatrix} F\\
 &=F^{-1} \begin{bmatrix}
(I_l +D) D&  \\
 & 0_{m \times m}C\\
 \end{bmatrix}F
 \end{align*}
  Notice that $(I_l +D)D=(D+D^2)=(D+I_l)$. Hence, 
  \begin{align*}
 QM&=F^{-1} \begin{bmatrix}
I_l+D &  \\
 & 0_{m \times m}\\
 \end{bmatrix}F=Q.
 \end{align*}
Using induction we see that $QM^{\theta}=QM M^{\theta -1}=QM^{\theta-1}=...=Q$. \\Thus, we get equation \eqref{mq_q}. \\
\item ${{(M}^\alpha +RQ)}^\theta\ ={\ M}^{\alpha\theta}+\mathrm{\Sigma Q}$.\\
We use induction on $\theta$. For $\theta=1$, the equation holds by taking $\Sigma=R$.\\
Now, suppose that ${{(M}^\alpha+RQ)}^\theta=M^{\alpha \theta}+\Sigma^\prime Q$, for some matrix $\Sigma^\prime\in\mathbb{F}_2^{n\times n}$.\\
Consider\[
  {{(M}^\alpha+RQ)}^{\theta+1}  \ ={{(M}^\alpha+RQ){(M}^\alpha+RQ)}^\theta \]
Hence, by the induction hypothesis 
\begin{align*}
(M^{\alpha}+RQ)^{\theta+1} &= (M^{\alpha}+RQ)(M^{\alpha\theta}+\Sigma^\prime Q) \\
  &=\ M^{\alpha(\theta+1)}+{M^{\alpha} \Sigma}^\prime Q+RQM^{\alpha\theta}+RQ\Sigma^\prime Q.
\end{align*}By part (a) we get $RQM^{\alpha\theta}=RQ$. Thus,
\begin{align*}
(M^{\alpha}+RQ)^{\theta+1}&=M^{\alpha(\theta+1)}+M^{\alpha}\Sigma^\prime Q  + RQ +RQ \Sigma^\prime Q\\
                    &=M^{\alpha(\theta+1)}+(M^{\alpha}\Sigma^\prime+R+RQ\Sigma^\prime)Q \\
                   &= M^{\alpha(\theta+1)}+\Sigma Q,
\end{align*}
where $\Sigma={M^\alpha\Sigma}^\prime+R+RQ\Sigma^\prime$. \\
Therefore, ${{(M}^\alpha +RQ)}^\theta\ ={\ M}^{\alpha\theta}+\mathrm{\Sigma Q}$,  for every positive integer $\theta$. \\
\item $(M^{\theta} + \Sigma _1 Q)(M^{\vartheta} + \Sigma _2 Q)=M^{\theta+\vartheta}+\Sigma _3 Q$.\vspace{6pt} \\
Consider 
 \[(M^{\theta} + \Sigma _1 Q)(M^{\vartheta} + \Sigma _2 Q)=M^{\theta+\vartheta}+M^{\theta}\Sigma _2 Q + \Sigma _1Q M^{\vartheta}+ 
 \Sigma _1 Q \Sigma _2 Q .\]
 From part (a), $\Sigma _1Q M^{\vartheta}$ $=$ $\Sigma _1Q$. Thus, 
 \begin{align*}
 (M^{\theta} + \Sigma _1 Q)(M^{\vartheta} + \Sigma _2 Q)
 &=M^{\theta+\vartheta}+M^{\theta}\Sigma _2 Q + \Sigma _1Q + 
 \Sigma _1 Q \Sigma _2 Q \\
 &=M^{\theta+\vartheta}+(M^{\theta}\Sigma _2  + \Sigma _1+ 
 \Sigma _1  \Sigma _2 )Q \\
 &=M^{\theta+\vartheta}+\Sigma _3 Q, 
 \end{align*}
 where $\Sigma _3=(M^{\theta}\Sigma _2  + \Sigma _1+ 
 \Sigma _1  \Sigma _2 $).
\end{enumerate}
\begin{flushright}
$\square$
\end{flushright}
\end{proof}
\section{New Hardness Assumptions}
\textsl{Decisional Diffie-Hellman (DDH)} assumption \cite{Boneh1998} is one of the standard hardness assumptions that are used for security proof in public-key cryptography. Let $\mathcal{G}$ be cyclic group with generator $g$ and order $q$ such that $q$ is prime, \textsl{DDH}  assumption in the group $\mathcal{G}$  states that it is hard to distinguish the distribution $(g,g^{\alpha}, g^{\beta},g^{\alpha \beta})$ from the distribution $(g,g^{\alpha}, g^{\beta},g^{\gamma})$, for uniformly random numbers $\alpha$, $\beta$, and $\gamma$ from $\mathbb{Z}_q$ with $\gamma \neq \alpha \beta$.   
 
Concretely, \textsl{DDH} assumption with respect to the general linear group $GL(n,2)$ is stated as follows. \\
\emph{It is hard distinguish between the two distributions
\[ (M,M^{\alpha},M^{\beta}, M^{\alpha \beta}) \text{ and } (M,M^{\alpha},M^{\beta},M^{\gamma}),\]
for uniformly random matrix $M \in GL(n,2)$  with multiplicative order $q$, such that $q$ is prime, and uniformly random numbers $\alpha$, $\beta$, and $\gamma$ form $\mathbb{Z}_q$ with $\gamma \ne \alpha\beta$}. \\

 \indent It is obvious that \textsl{DDH} assumption relies on the hardness of the \textsl{Discrete Logarithm Problem (DLP)} in the relevant group in which it is defined. Accordingly, the \textsl{DDH} assumption in $GL(n,2)$ relies on the hardness of the \textsl{DLP} in this group. However, there is polynomial-time reduction of the \textsl{DLP}  in $GL(n,2)$ to \textsl{DLP}  in some small extension fields of $\mathbb{F}_2$ as shown in \cite{menezes1997discrete}. This renders the \textsl{DDH} assumption in $GL(n,2)$ invalid against Shor's algorithm \cite{shor1994algorithms}  for the \textsl{DLP}. \\
 \indent To overcome this problem we propose involving random errors in the Diffie-Hellman quadruple 
of matrices $(M,M^{\alpha},M^{\beta}, M^{\alpha \beta})$ so as to destroy the discrete logarithm relations among its components; that is to say, to destroy the discrete logarithm relations $(M,M^{\alpha})$, $(M,M^{\beta})$, $(M^{\alpha},M^{\alpha\beta})$, and ($M^{\beta}, M^{\alpha\beta})$. Thereby we prevent the quadruple from being distinguishable from the random by solving some \textsl{DLP}, and thus we get new hardness assumption which bears resemblance to \textsl{DDH} assumption and inherits its well-known hardness, but unlike the standard \textsl{DDH} assumption, the new assumption is secure against quantum attack using Shor's algorithm, since it is no longer a \textsl{DLP}-based assumption. \\ 
\indent More precisely, consider the Diffie-Hellman quadruple $(M,M^{\alpha},M^{\beta}, M^{\alpha \beta})$  and let $E_0$, $E_1$, and $E_3$ be matrices sampled uniformly  at random from some \emph{Error Distribution} $ \mathcal{E} \subset \mathbb{F}_2 ^{n \times n}$, we destroy the discrete logarithm relations among elements of this quadruple  by taking bitwise XOR of each one of the  matrices $M$, $M^{\alpha}$ and $M^{\beta}$ , respectively, with each of the matrices $E_0$, $E_1$, and $E_3$   to produce new quadruple   
 \[  \mathcal{Q}_{M,\alpha\beta}=(M\oplus E_0,M^{\alpha}\oplus E_1,M^{\beta} \oplus E_2, M^{\alpha \beta}).\]
 The quadruple $\mathcal{Q}_{M,\alpha\beta}$ does not, any more, contain \textsl{DLP} among its components. We call the  quadruple $\mathcal{Q}_{M,\alpha\beta}$ \textit{Noisy  Diffie-Hellman Quadruple} \textit{(N-DHQ)} in $GL(n,2)$, and we assume that this quadruple is computationally indistinguishable from the random corresponding counterpart \textit{non-Dffie-Hellman} quadruple of matrices with noise. 
 
 \subsection{Definitions}
 \subsubsection{Algorithm: \textsl{$Gen_{dh-we}(n,b)$}.} Consider the matrix $M$ as defined in equation (\ref{M}). Let $Gen_{dh-we}$ be probabilistic polynomial-time algorithm that on input ${(n,b \in \{0,1\})}$ returns $\mathcal{D}_{\delta} =(A,B,S,Y,v_{\delta} = M^{\delta}_j)$, where
\[ A = M \oplus R_0Q, B =M^{\alpha} \oplus R_1 Q, Y= M^{\beta} \oplus R_2Q,\text{ and } S=RQ,\]
for uniformly random matrices $R$, $R_0$, $R_1$, and $R_2$ from $\mathbb{F}_2^{ n \times n}$ and  uniformly random numbers $\alpha$, $\beta$ and $\delta$ from $\mathbb{Z}_{\phi}$, such that $\delta = b \alpha\beta \mod \phi + (1-b)\gamma$, $\gamma \in \mathbb{Z}_{\phi}$ (thus, $v_{\delta}$ equals either $M^{\alpha\beta}_j$ or $M^{\gamma} _j$), and $\phi$ is the multiplicative order of $M$. The subscript $_j$ indicates the $j$th column of the matrix. 
 
\noindent \textbf{Note:} The matrix $S$ is independent of the numbers $\alpha$, $\beta$, $\gamma$ and the bit $b$.  
\subsubsection{Algorithm: \textsl{$Gen_{rsa-r}(n,b)$}.} Consider the matrix $M$ as defined in equation (\ref{M}). Let $Gen_{rsa-r}$ be probabilistic polynomial-time algorithm that on input ${(n,b \in \{0,1\})}$ returns $\mathcal{D}_{\delta,\phi} =(A,B,e,Y,v_{\delta} = M^{\delta}_j)$, where
\[ A = M \oplus R_0Q, B =M^{\alpha} \oplus R_1 Q,\text{ and } Y= M^{\beta} \oplus R_2Q \]
for uniformly random matrices  $R_0$, $R_1$, and $R_2$ from $\mathbb{F}_2^{ n \times n}$,  uniformly random numbers $\alpha$, $\beta$, $\delta$ from $\mathbb{Z}_{\phi}$, and $e \in \mathbb{Z}^*_{\phi}$, such that $\delta = b e^{-1} \mod \phi  + (1-b)\gamma$, $\gamma \in \mathbb{Z}_{\phi}$ (thus, $v_{\delta}$ equals either $M^{d\beta}_j$ or $M^{\gamma} _j$), where $ d= e^{-1} \mod \phi$ and $\phi$ is the multiplicative order of $M$. \\

Notice that in both of the two algorithms the matrices  $R_0Q$, $R_1Q$, $R_2Q$, and $S=RQ$ belongs to the \textsl{Error Distribution}  
 \[ \mathcal{E}=\{EQ = E(I_n \oplus M^{r}),E \in \mathbb{F}_2 ^{n \times n} \}. \]
 And, the $j$th column of each of one these matrices is zero. That is, 
 \[ ({R_0 Q})_j =({R_1 Q})_j={(R_2 Q})_j=(S_j =({RQ})_j) =0,\]
 because the $j$th column of the matrix $Q$ is zero (i.e., because $Q_j$ = 0). 
 \begin{remark}
 \label{label=clmsrnd}
Almost all columns of the matrices $A = M \oplus R_0Q$, $B = M^{\alpha} \oplus R_1Q$, and $Y = M^{\beta} \oplus R_2Q$ are different from their corresponding counterparts of $M$, $M^{\alpha}$, and $M^{\beta}$, respectively, except for the $j$th column in each matrix, since the $j$th column of the matrix $Q$ is zero column and hence  the $j$th column of $EQ$, for every matrix $E$, is zero column too. In other words, 
 if $A=[A_1 \cdots A_m ], B= [B_1 \cdots B_n], \dots \text{and so on for rest the of matrices }$ $ M,Y, M^{\alpha}, \text{ and } M^{\beta}$, then 
\[
A_k = 
\left\{
 \begin{array} {lr} 
 A_k \in \mathbb{F}_2 ^n, & \text{ if } k \neq j \\
 M_j, & \text{ if } k=j
\end{array} 
\right\}
, 
B_k = 
\left\{
 \begin{array} {lr} 
 B_k \in \mathbb{F}_2 ^n, & \text{ if } k \neq j \\
 M^{\alpha}_j, & \text{ if } k=j 
\end{array} 
\right\}, \]
 \[ \text{ and }Y_k = 
\left\{
 \begin{array} {lr} 
 Y_k \in \mathbb{F}_2 ^n, & \text{ if } k \neq j \\
 M^{\beta}_j, & \text{ if } k=j 
\end{array} 
\right\}
.\]
\end{remark}
\begin{definition} \textsl{(DDH Assumption with Errors (DDH-WE))} Let $\mathcal{A}$ be probabilistic polynomial-time adversary who has an oracle access to the output  $\mathcal{D_{\delta}}$ of the algorithm $Gen_{dh-we}$ on the input $(n, b), b \in \{0,1\}$. We define advantage of $\mathcal{A}$ in figuring out the random bit $b$ as follows. 
\[ Adv_{\mathcal{D_{\delta}}}(\mathcal{A})= |1-2*P[b'=b: b \xleftarrow{R} \{0,1\};\mathcal{D_{\delta}} \gets Gen_{dh-we}(n,b); b' \gets \mathcal{A}(\mathcal{D_{\delta}}) ]| \]
where the probability is taken over $\mathcal{D_{\delta}}$ and coin tosses of the adversary $\mathcal{A}$.\\
The \textsl{DDH-WE} assumption states that advantage of the adversary $\mathcal{A}$ in guessing   value the of the random bit $b$ is a negligible function of $n$. In other words, 
\[ Adv_{\mathcal{D_{\delta}}}(\mathcal{A})\approx neg(n). \]
 \end{definition} 
\begin{definition} \textsl{(Decisional \textsl{RSA}-Resemble (DRSA-R) Assumption  )} Let $\mathcal{A}$ be probabilistic polynomial-time adversary who has an oracle access to the output  $\mathcal{D_{\delta,\phi}}$ of the algorithm $Gen_{rsa-r}$ on the input $(n, b), b \in \{0,1\}$. We define advantage of $\mathcal{A}$ in figuring out the random bit $b$ as follows. 
\[ Adv_{\mathcal{D_{\delta,\phi}}}(\mathcal{A})= |1-2*P[b'=b: b \xleftarrow{R} \{0,1\};\mathcal{D_{\delta,\phi}} \gets Gen_{rsa-r}(n,b); b' \gets \mathcal{A}(\mathcal{D_{\delta,\phi}}) ]| \]
where the probability is taken over $\mathcal{D_{\delta,\phi}}$ and coin tosses of the adversary $\mathcal{A}$.\\
The \textsl{(DRSA-R)} assumption states that advantage of the adversary $\mathcal{A}$ in guessing   value the of the random bit $b$ is a negligible function of $n$. In other words, 
\[ Adv_{\mathcal{D_{\delta,\phi}}}(\mathcal{A})\approx neg(n). \]
 \end{definition}
 
 \begin{definition}
A key-agreement is an IND-RND secure if it is hard to distinguish the shared secret key which results from it from the random.
\end{definition}
\subsection{Why the New Assumptions Should Hold?}
\textbf{The first assumption:} As we have seen,  the \textsl{DDH-WE} is obtained by introducing random noise into \textsl{DH} quadruple in $GL(n,2)$. Thus, if \textsl{DDH} assumption in $GL(n,2)$ is hard, then  it remains hard after adding random noise into the quadruple by which it is defined. Therefore, the \textsl{DDH-WE} assumption is, at least, as hard as \textsl{DDH} assumption in $GL(n,2)$. Furthermore, unlike the  \textsl{DDH} assumption which is subject to \textsl{DLP}-based attacks, the new assumption is, as far as we can see, secure against \textsl{DLP}-based attacks including Shor's algorithm for the \textsl{DLP}.\\
\textbf{The second assumption:} \textsl{DRSA-R} bears resemblance to the \textsl{DDH-WE} assumption, and by transitivity it resembles the \textsl{DDH} in $GL(n,2)$. It consist of the noisy triple $M$, $M^{\alpha}$, and $M^{\beta}$, in addition to the column vector $M^{\delta}_j$. But unlike \textsl{DDH} the number $\delta$ does not take the value $\alpha\beta$; instead $\delta = d\beta$ where $d$ is multiplicative inverse of given number $e$ modulo a secret number $\phi$ which is multiplicative order of the matrix $M$ of which only noisy powers are available. Assuming that it is hard to find $\phi$ (and thus it is hard to find $d$), the \textsl{DRSA-R} is similar to \textsl{DDH} with noise in $GL(n,2)$. Therefore, we assume it is hard. Also, \textsl{DRSA-R} has similarity with $RSA$ problem; that is, given $Y=X^{e}$ and $e$, find $X$, and also the problem: given the column $Y^{d'}_j$  and the number $e$, decide if $d'=d$. Luckily, whilst \textsl{RSA} problem is breakable using Shor's algorithm for integer factoring, no similar attack on $(e,d)$ in the new assumption. 
\section{Diffie-Hellman Key-Agreement with Errors (\textsl{DH-WE})}
The \textsl{DDH-WE} can be used as a basis for the following key-agreement algorithm between the two parties Alice and Bob. In what follows give detailed description of the \textsl{DH-WE} key-agreement, prove its correctness and its IND-RDN security under \textsl{DDH-WE} assumption. 
\begin{algorithm}[H]
\caption{\textsl{DH-WE} Key-agreement}
\begin{algorithmic}[1]
\item \textit{Alice performs the following steps}:
\begin{enumerate}[label=(\alph*)]
\item  Generate secret random matrix $M$ of order $\phi=2r$ as described in equation \eqref{M}.  Compute $Q={(I_n \oplus M}^r)$. 
\item Generate secret random matrices $R$, $R_0$, and $R_1$.  Select  secret random number $\alpha$ such that $1< \alpha < r$. Compute \[A = M \oplus R_0Q \text{, } B=M^{\alpha}\oplus R_1Q \text{, and }  S=RQ, \] such that ${det{\left(A\right)}\neq 0}$.
\Comment{Thus, $A = M \oplus E_0$ and $B = M^{\alpha} \oplus E_1$, where $E_0 = R_0 Q$ and $E_1 = R_1 Q$.}
\item Send $(A, B, S)$ to Bob. 
\end{enumerate}
\item \textit{Bob receives $(A, B, S)$ and performs the following steps}: 

\begin{enumerate}[label=(\alph*)]
\item Select secret random number $\beta$ such that $1 < \beta < 2^{n}-1$. 
 \\ Compute the secret key $k_B = B^{\beta}_j$ -- the $j$th column of the matrix $B^{\beta}$. 
 
 \Comment{Notice that Alice communicates the number $j$ to Bob through the matrix $S$, where $j$ is the position of the zero column in $S$.}
\item  Generate secret random matrix $R_2$. Compute \[ {Y=A^{\beta} \oplus R_2S}.\]
 \item Send $Y$ to Alice.
\Comment{Indeed $Y=A^{\beta} \oplus  R_2 S = M^{\beta} \oplus E_2 $, for some unknown matrix $E_2$.}

\end{enumerate}
\item Alice computes the secret key $k_A = Y^{\alpha}_j$ -- the $j$th column of the matrix $Y^{\alpha}$. 
\end{algorithmic}
\end{algorithm}
The shared secret key is \[k_A=Y^{\alpha}_j=B^{\beta}_j=k_B.\] 
\indent The key-agreement is specified by  the \emph{tuple} $\Pi_{M,\alpha\beta}$ = $(A,B,S, Y,Y^{\alpha}_j)$. That is, the key-agreement as an exchange of messages between two parties via public communication channel is captured by this tuple. Notice that one can specify the key-agreement as $\Pi_{M,\alpha\beta}$ = $(A,Y,S,B,B^{\beta}_j)$ instead. It would essentially be the same specification. 
\subsection{Correctness of the Key-agreement}
We see by Theorem \ref{thrm1} that for every matrix $M$ defined as in equation (\ref{M}), matrix $Q=( I_n \oplus M^{r})$, two integers $\alpha$ and $\beta$, and any random matrix $R\in \mathbb{F}^{n \times n} _2$, the following equation holds.
\begin{equation}
\label{Mab}
 (M^{\alpha} \oplus RQ)^{\beta} = M^{\alpha\beta}\oplus \Sigma Q, 
\end{equation}
where $\Sigma$ is some unknown matrix in $\mathbb{F}^{n \times n} _2$.\\

Since $Y= A^{\beta}\oplus R_2S$, where $A = M \oplus R_0Q$ and $S= RQ$, it follows that 
\begin{align*}
Y &=(M \oplus R_0Q)^{\beta} \oplus R_2RQ \\
&=M^{\beta} \oplus \Sigma Q \oplus R_2RQ \hspace{12 pt} (\text{using equation (\ref{Mab})})\\
&=  M^{\beta} \oplus \Sigma_1 Q,  \text{ where } \Sigma_1=(\Sigma \oplus R_2R).
\end{align*}
Next, since $Y= M^{\beta} \oplus \Sigma_1 Q$ and  $B = M^{\alpha}\oplus R_1Q$, we get, by equation (\ref{Mab}),
\[ Y^{\alpha} = M^{\alpha \beta} \oplus \Sigma_2Q \text{ and } B^{\beta}= M^{\alpha\beta} \oplus  \Sigma_3 Q, \]
for some unknown matrices $\Sigma_2,\Sigma_3 \in \mathbb{F}^{n \times n} _2$.\\
 However, since $Q_j=0$, which implies that $(EQ)_j=0$ for every matrix $E \in \mathbb{F}^{n \times n} _2$ (see Remark \ref{label=clmsrnd}) and hence $(\Sigma_2Q)_j =(\Sigma_3Q)_j = 0$, it follows that 
\[ (k_A= Y^{\alpha}_j)=(k_B= B^{\beta}_j)=M^{\alpha\beta}_j). \] 
\subsection{Security Proof}
\begin{definition}
The key-agreement $\Pi_{M,\alpha\beta}$ is said to be IND-RND secure if it is hard to distinguish the shared secret key $k=Y^{\alpha}_j$ from the random vector $k^* = Y^{\gamma}_j$. More precisely, the key-agreement $\Pi_{M,\alpha\beta}$ is said to be IND-RND secure if it is hard to distinguish 
 \[\Pi_{M,\alpha\beta}=(A,B,S,Y,Y^{\alpha}_j)\]  from  \[ \Pi_{M,\gamma}=(A,B,S, Y,Y^{\gamma}_j) \text{ with } \gamma\beta \neq \alpha\beta. \]
\end{definition}

\begin{theorem}
\label{proof1}
The key-agreement $\Pi_{M,\alpha\beta}$ is IND-RND under DDH-WE assumption. 
\end{theorem}

\begin{proof} 
Since, \[ 
\Pi_{M,\alpha\beta} \equiv \mathcal{D}_{\alpha\beta} = (A,B,S,Y,v_{\alpha\beta} = M^{\alpha\beta}_j) \leftarrow Gen_{dh-we}(n,1),  \]
and 
\[ 
\Pi_{M,\gamma}\equiv \mathcal{D}_{\gamma} = (A,B,S,Y,v_{\gamma} = M^{\gamma}_j) \leftarrow Gen_{dh-we}(n,0),  \]
the theorem holds directly by definition of the \textit{DDH-WE} assumption. 
\begin{flushright}
$\square$
\end{flushright}
\end{proof}
\section{\textsl{RSA}-Resemble  Key-agreement}
 The \textsl{RSA}-Resemble key-agreement scheme is constructed using almost the same components of the \textsl{DH-WE} key-agreement scheme. The new key-agreement is an IND-RND secure under \textsl{DRSA-R} assumption.
\begin{algorithm}[H]
\caption{\textsl{RSA}-Resemble Key-agreement}
\begin{algorithmic}[1]
\item \textit{Alice performs the following steps}:
\begin{enumerate}[label=(\alph*)]
\item  Generate secret random matrix $M$ of order $\phi=2r$ as described in equation \eqref{M}.  Compute $Q={(I_n \oplus M}^r)$. 
\item Select secret random  number $e \in {\mathbb{Z}}_{\phi} ^{*}$. Compute $d$ = $e^{-1} \mod \phi$.\vspace{1pt}
\item Select  secret random number $\alpha$ such that $1< \alpha < r$. Generate secret random matrices $R_0$ and $R_1$. Compute \[A = M \oplus R_0Q \text{ and } B=M^{\alpha} \oplus R_1Q\] such that ${det{\left(B\right)}=0}$.\vspace{1pt}
\item Send  $(A, B,e,j)$ to Bob and keep $d$ privately.
\end{enumerate}
\item \textit{Upon receiving  $(A, B,e,j)$ Bob performs the following steps}: 
\begin{enumerate}[label=(\alph*)]
\item  Select secret random nonnegative integers  $\theta$ and $\vartheta$ from the set $[0,2^{n/2}]$. Compute{ $X=A^{\theta}$$B^{\vartheta}$ and set $k_B=X_j$} -- the $j$th column of the matrix $X$.
\item Compute  $Y=X^e  (= M^{\beta} \oplus E_2$, for some unknown matrix $E_2$), and send $Y$ to Alice. 
\Comment{Notice that $\beta = e(\theta + \alpha \vartheta) \mod \phi $ as we will see in equation \eqref{Y_M_Beta}.}
\end{enumerate}
\item Alice computes $k_A=Y^d_j$, the $j$th column of the matrix $Y^d$ using the secret number $d$.
\end{algorithmic}
\end{algorithm}
The shared secret key is $k=(k_A=Y^{d}_j)=(k_B=X_j)$. The \textsl{RSA}-Resemble key-agreement scheme is specified by the \textit{tuple} \[\Pi_{M,e,d}= (A,B,e,Y,k).\]
\subsection{Correctness of the Key-agreement}
We show that both Alice and Bob share the same secret key. That is, $k_A$ = $k_B$. 
\subsubsection{Bob's Key:}
\begin{align*}
   X&=A^{\theta} B^{\vartheta}\\
   &=(M +R_0 Q)^{\theta}(M^{\alpha}+R_1 Q)^{\vartheta} \\
   &=(M^{\theta}+\Sigma_1 Q)(M^{\alpha \vartheta}+\Sigma_2 Q) \text{\hspace{12pt}(using equation \eqref{m_alpha_rq_rq}}\\
   &= (M^{\theta+\alpha \vartheta}+\Sigma_3 Q) \text{\hspace{12pt}(using equation \eqref{m_alpha_beta_rq_rq})}  
\end{align*}
\indent Since the $j$th column of the matrix $Q$ is zero, 
\[k_B=X_j=M^{ \theta + \alpha \vartheta}_j.\]
In other words, the key $k_B$ is the $j$th column of the matrix $X$, which is equal to the $j$th column of the matrix $M^{ \theta+\alpha\vartheta}$.  
\subsubsection{Alice's Key:} 
\begin{align*}
Y^d&=(X^e)^d  \\
&=(A^{\theta} B^{\vartheta})^{ed} \\
   &=((M + R_0 Q)^{\theta}(M^{\alpha}+R_1 Q)^{\vartheta})^{ed} \\
   &=((M^{ \theta}+\Sigma_1Q )( M^{\alpha \theta} +\Sigma _2 Q))^{ed} \hspace{12pt} \text{(using equation \eqref{m_alpha_rq_rq})}\\
&=(M^{(\theta+\alpha\vartheta)}+\Sigma_3 Q)^{ed}
\text{\hspace{12pt}(using equation \eqref{m_alpha_beta_rq_rq})}\\
&=(M^{ed(\theta+\alpha\vartheta)}+\Sigma_4 Q) \text{\hspace{12pt}(using equation \eqref{m_alpha_rq_rq})}\\
&=(M^{(\theta+\alpha \vartheta)}+\Sigma_4 Q) \text{\hspace{12pt}(since $ed=1\mod \phi$)}
\end{align*}
\indent Now, we know that the $j$th column of the matrix $Q$ is zero, therefore 
\[k_A=Y^{d}_j=M^{\theta+ \alpha  \vartheta}_j=k_B.\]
Notice that 
\begin{equation}
\label{Y_M_Beta}
Y = M^{e(\theta+\alpha\vartheta)} + \Sigma_4 ^{'}Q = M^{\beta} + E_2, 
\end{equation}
where $\beta = e(\theta+\alpha\vartheta)$ and $E_2 = \Sigma_4 ^{'} Q $, for some matrix $\Sigma_4 ^{'}$ with $\Sigma_4 ^{'} \neq
\Sigma_4$.
\begin{flushright}
$\square$
\end{flushright}
\subsection{Security of the \textsl{RSA}-Resemble Key-agreement}
 \begin{theorem}
\label{label=proof1}
The key-agreement $\Pi_{M,e,d}$ is IND-RND under DRSA-R assumption. 
\end{theorem}
The proof of this theorem is almost identical to the proof Theorem \ref{proof1}, since  
$\Pi_{M,e,d} \gets Gen_{rsa-r}(n,1)$ so there is no need to repeat the same proof. The theorem holds directly by the definition of the \textsl{DRSA-R} assumption.
\subsection{Notes About  One-wayness of the Operation $Y \gets X^e$}
\indent As mentioned earlier, we assume that it is hard to find the private number $d$, since the  matrix $M$ is secret and thus its multiplicative order $\phi$ is also secret. Furthermore, the following theorem shows that  it is unlikely there exists number $d'$ such that $Y^{d'}=X$. 
\begin{theorem}
For any integer $e>1$, it is unlikely there exists nonnegative integer $d'$ that satisfies $Y^{d'}= X$. 
\end{theorem}
\begin{proof}
We have $\det(X)=0$, since $X=A^{\theta}B^{\vartheta}$ and $\det(B)=0$. Therefore, the matrix $X$ has no multiplicative order. In other words, there is no integer $s$ such that $X^{s}=I_n$. \\
\indent Now, let $s$ be the smallest positive integer such that 
\[
X^s = X^t, \hspace{4pt} 0<t<s.
\]
For every two positive integers $e$ and $d'$ we have
\[ed'=\lambda s+r', \hspace{4pt}\text{where} \hspace{4pt} \lambda  \geq 0 \hspace{4pt} \text{and} \hspace{4pt} 0 \leq r'<s. \]
Thus, 
\begin{equation}
\label{thmeq1}
Y^{d'} = X^{ed'} = X^{\lambda s+r'} = X^{\lambda t+r'}
\end{equation}
From equation \eqref{thmeq1} we see that \[ Y^{d'}=X^{ed'}=X \] if and only if $ed'=s$ and $t=1$. In other words, $\lambda=1$ and $r'=0$.\vspace{6pt}\\
The probability of such number $d'$ is a probability that $e$ divides $s$ and $t=1$. That is,  
\[P[e|s \hspace{4pt}\text{and} \hspace{4pt} t=1].\] 
\noindent However, the number $e$ is random independently  selected and both the  numbers $s$ and $t$ are also random  with $1 \leq t < s$. Therefore, the likelihood of existence of the number $d'$ is negligibly  small when the size of the matrix $X$ is large.
\begin{flushright}
$\square$
\end{flushright}
\end{proof}
\indent In addition to unlikelihood of existence of the number $d'$, it is also hard to be figured out even when it exists, because the matrix $X$ is secret and hence the number $s$ is, anyway, unknown.
\begin{remark}
In particular, $Y^{d}\neq X$, where $d$ is the multiplicative inverse of $e$ modulo $\phi$. Actually, $Y^{d} = M^{ \theta +\alpha \vartheta}+\Sigma _4 Q$, for some unknown matrix $\Sigma _4$ whereas  $X=M^{ \theta +\alpha \vartheta}+\Sigma _3 Q$, for another unknown matrix $\Sigma _3$, as we have seen in Section 4.2. 
\end{remark}

\subsection{Brute Force Attacks}
Security against brute force attack in both of the two schemes depends on the size of the multiplicative order $\phi$ of the matrix $M$. Precisely, the number $\phi= 2r$ depends on the number $r$ which is the multiplicative order of the matrix $C$, (see equation \eqref{M}). \\
\indent In the \textsl{DH-WE} scheme we can maximize $\phi$ up to $\Theta\left(2^m\right)$, since the size of the matrix $C$ is $m \times m$ and thus its multiplicative order $r$ can be up to $2^{m}-1$. Therefore, exhaustive search for each of the secret  exponents $\alpha$ and $\beta$ is of order $\Theta\left(2^m\right)$. Exhaustive search for the shared secret key $k_A = k_B \in \mathbb{F}^{n} _2$  is of order ${\Theta}\left(2^n\right)$. Recall that $n = m+l$, thus in order to maximize the multiplicative order $\phi$ we choose the number $l$ to be small number while $m$ is large. For example, $l=10$ and $m = 118$ and thus $n=128$. \\ 
\indent Regarding the \textsl{RSA}-Resemble scheme, where the  multiplicative order $\phi$ of $M$ is secret, exhaustive search attack on the private key $d \in \mathbb{Z}_{\phi} ^{*}$ depends on the size  of $\phi$ which is random number between 1 and $2^m -1$. However, we should mention that more investigation on the density  of the set of all possible values of $\phi$ is required. Brute force attack on the secret key $k_B = X_j=A^{\theta}B^{\vartheta}_j$, where ${\theta}$,${\vartheta}$ $\in \{0,1\}^{n/2}$, has the same complexity of the exhaustive search for the key $k_B= k_A \in \mathbb{F}_2 ^n$ (i.e., $\Theta\left(2^n\right)$).   
\section{Conclusion}
Security of each of the \textsl{DH} key-agreement and the \textsl{RSA} cryptosystem has been studied for long time. Security of each of the two schemes is, respectively, characterized by (a) hardness of the \textsl{DLP} and (b) hardness of obtaining the multiplicative inverse a given number $e$ modulo secret inverse $\phi$. The two proposed schemes have these two characteristics, and thus they, in some sense, inherit security against key-recovery of each of the \textsl{DH} key-agreement and the \textsl{RSA},  respectively, without however being vulnerable to Shor's algorithms attack like the two classical algorithms do. In particular, with regards to security of \textsl{DH-WE}, we know from history of \textsl{DH} key-agreement that it is hard to find any of the secret exponents $\alpha$ and $\beta$ without solving a \textsl{DLP}, and also from history of \textsl{RSA} algorithm we know that given the number $e$ it is hard to find its inverse $d$ modulo secret number $\phi$ without finding the number $\phi$ itself.  
What we do in this paper is precluding Shor's algorithms attacks by corrupting the \textsl{DLP} (i.e., turning it into a \textsl{DLP} with errors) in the first scheme and by getting rid of the modulus $n = pq$ used by \textsl{RSA} in the second scheme.\\
Furthermore, the two schemes have been proved to be secure in the stronger  notion of indistinguihability of the secret key from random  under new family hardness assumptions which is connected to the standard \textit{DDH} assumption in $GL(n,2)$. The underlying security assumption for both of the two schemes is rooted in the hardness of learning information from noisy binary matrices. Thus, security assumption of the two proposed schemes has in-commons with \textit{code-based} as well as \textit{Learning Parity with Errors (LPN)} \cite{pietrzak2012cryptography} and  \textit{Learning with Errors (LWE)}\cite{regev2009lattices} hardness assumptions which are used as basis for some quantum-based cryptographic algorithms. Further analysis can be done in this regard.\\
\indent Future work will include implementation, performance analysis, and comparison with some existing quantum-based algorithms. 

\def\bibfont{\small}
\bibliographystyle{splncs04}
\bibliography{DH-RSA.bbl}
\end{document}